\documentclass[twocolumn]{jpsj2}
\usepackage{graphicx}% Include figure files
\usepackage{dcolumn}% Align table columns on decimal point
\usepackage{bm}% bold math

\newcommand{\ep}{\varepsilon}
\renewcommand{\Im}{\mathop{\mathrm{Im}}}
\renewcommand{\Re}{\mathop{\mathrm{Re}}}

\def\runauthor{O. \'Ujs\'aghy and A. Zawadowski}

\begin{document}

\title{Kondo Effect on Mesoscopic Scale (Review)}

\author{O. \'Ujs\'aghy$^1$ and A. Zawadowski$^{1,2}$}
\inst{$^{1}$Budapest University of Technology and Economics,
Institute of Physics and Research group Theory of Condensed Matter of 
Hungarian Academy of Sciences
H-1521 Budapest, Hungary \\
$^{2}$Research Institute for Solid State Physics, POB 49, H-1525
Budapest, Hungary}

\recdate{\today}

\abst{Following the discovery of the Kondo effect the bulk transport and
  magnetic behavior of the dilute magnetic alloys have been successfully
  described. In the last fifteen years new directions have been developed as
  the study of the systems of reduced dimensions and the artificial atoms so
  called quantum dots. In this review the first subject is reviewed starting
  with the scanning tunneling microscope (STM) study of a single magnetic
  impurity. The next subject is the reduction of the amplitude of the Kondo
  effect in samples of reduced dimension which was explained by the surface
  magnetic anisotropy which blocks the motion of the integer spin nearby the
  surface. The electron dephasing and energy relaxation experiments are
  discussed with the possible explanation including the surface anisotropy,
  where the situation in cases of integer and half-integer spins is very
  different. Finally, the present situation of the theory of dynamical
  structural defects is briefly presented which may lead to two-channel Kondo
  behavior.}
 
\kword{Kondo effect,Kondo resonance,size dependence,mesoscopic
  samples,spin-orbit-induced surface anisotropy,dephasing,energy
  relaxation,dynamical defects}

\maketitle 

\section{Introduction}

Following the pioneering work of Kondo \cite{Kondo} in the first decades the
properties of bulk dilute magnetic alloy systems have been successfully
studied. In the last fifteen years, however, the interest has turned to the
systems of reduced dimensions and the single impurity problems. The latter is
also represented by the very rapidly developing fields of quantum dots, where
Kondo like impurities are manufactured. The goal of the present review to have
an overlook over the most interesting problems of the first kind, and
especially, the connection between the phase coherence and energy relaxation,
the surface anisotropy and the nature of the Kondo scatterers will be
emphasized.  The limited length of this review does not allow completeness,
but it is restricted to some of the most important moments and phenomena.

\section{Kondo resonance in the electron density of states measured by
  STM}

\begin{figure}
\begin{center}
\includegraphics[height=4cm]{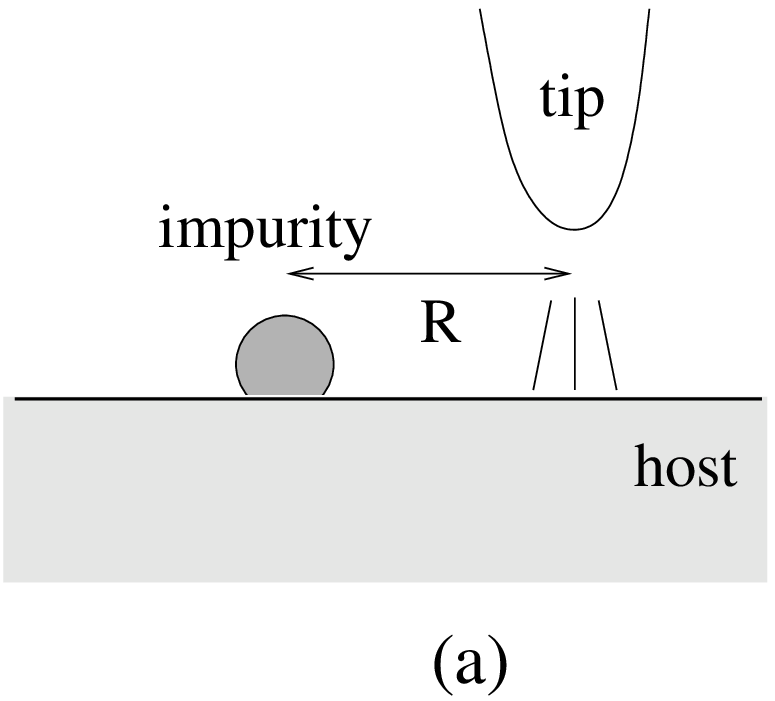}\vspace{1em}
\noindent\includegraphics[height=5cm]{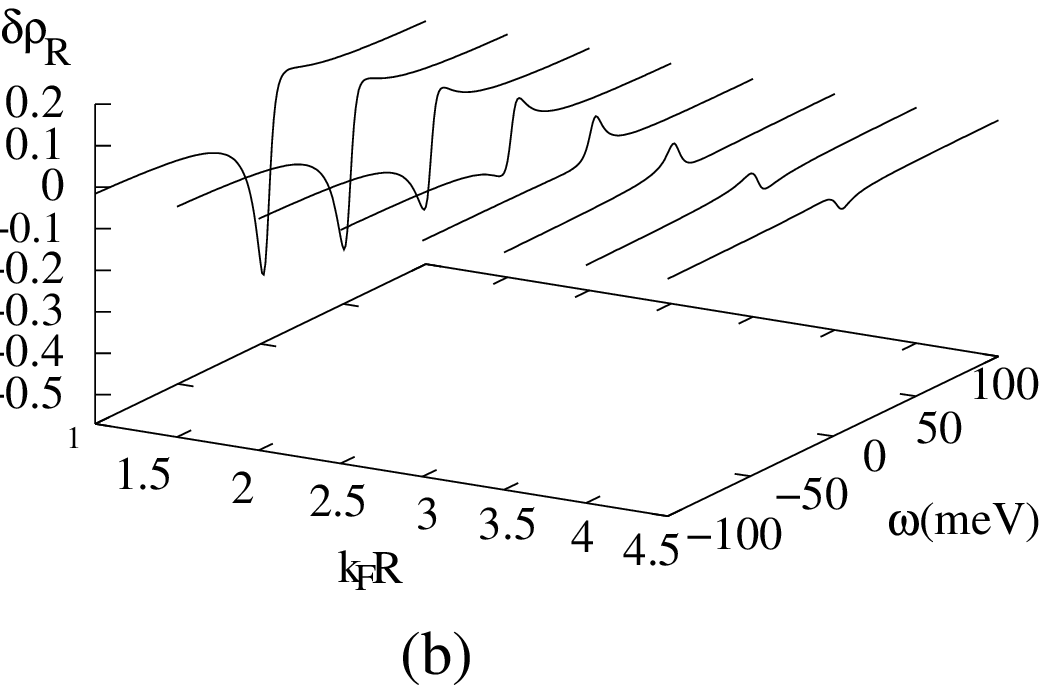}
\caption{(a) Schematic setup of the STM measurements.
  (b) Qualitative dependence of the calculated line shape of the tunneling DOS
  on the distance of the tip from the impurity using the first part of
  Eq.~(\ref{roveg}) and the 3-dimensional Au bulk Green's functions. The
  results obtained by using the 2-dimensional free-electron like Au(111)
  surface band \cite{Chen} are quite similar \cite{UKSZZ}.}
\label{fstm}
\end{center}
\end{figure}

In the original Kondo model based on the Kondo Hamiltonian the magnetic
impurity is represented as a localized spin. The Kondo resonance is around the
Fermi energy and frequently called Abrikosov-Suhl resonance. The more
appropriate model is, however, the Anderson model, where the d-levels with
finite width, $\Delta$, are also included, and they are split due to the
on-site Coulomb interaction. It was conjectured by Gr\"uner and one of the
authors \cite{GZ} and first calculated by Yamada \cite{Yamada} that the Kondo
resonance is superimposed on the atomic d-level structure in the d-level
density of states. Since that time it has been a great challenge to measure
that directly. It was shown \cite{MZ} that the conduction electron density is
suppressed nearby the impurity because the strong scattering by the impurity
results in interference between the incoming and outgoing waves and that
depression in the local density of states (LDOS) is the largest around the
Fermi energy in the range of the Kondo temperature \cite{MZ}. That time a
local probe of the density of states was not available. It was suggested,
however, that a dilute layer of magnetic impurities has similar effect. That
could be measured by placing that layer nearby the barrier in one of the
electrodes of a metal-oxide-metal tunnel junction, where the resonance occurs
as a zero bias resistivity maximum \cite{44}. Recently, by using scanning
tunneling microscope (STM) the direct probe became possible. Concerning the
experimental observation an unexpected difficulty appeared, namely the surface
anisotropy for integer spins to be discussed later. The Mn impurity has very
low Kondo temperature in Au, Ag and Cu, thus too high resolution would be
required. The Fe and Cr impurities do not move freely because of having $S=2$
which was not realized at the time of the first experiments, but were examined
later in a systematic study of different transition-metal impurities on Au
surfaces \cite{Jamneala}. The successful experiments were performed with
single Ce atoms on Ag \cite{Li} as well as with single Co atoms on Au
\cite{Madhavan} and Cu \cite{Manoharan} surfaces by measuring the I-V
characteristics of the tunneling current through the tip of a STM placed close
to the surface and at a small distance $R$ from the magnetic atom (see
Fig.~\ref{fstm} (a)).
In these experiments the electron tunnels from the tip into the metal, travels
to the impurity and after scattering goes back to the tip, resulting in an
interference between the unperturbed and scattered electron waves. There is
also a possibility that the electrons tunnel from the tip directly into the d-
or f-level of the magnetic impurity.  However, the tunneling rate for the
latter process is probably very small, especially for f-levels, which are
deeply inside the atom. To explain the observed narrow asymmetric Fano
resonance \cite{Fano} structure in the electronic LDOS the first proposed
theory \cite{Madhavan,Schiller} took into account both processes. However, as
it has been shown in Ref.~\cite{UKSZZ}, the Fano resonance can develop even if
one neglects the direct tunneling to the impurity. This assumption was
supported also by the ``quantum mirage experiments'' \cite{Manoharan} where an
elliptical corral was built from magnetic or non-magnetic atoms on a metal
surface with a magnetic impurity in one of the foci's and an attenuated Fano
resonance was measured in the other foci as well.  In that case the role of
the direct tunneling is obviously negligible. That assumption was also
justified later for the experimental situation by a microscopic theory of
tunneling into a single impurity on a metal surface by Gadzuk et al.
\cite{Gadzuk}.

In the theory of Ref.~\cite{UKSZZ} the physics is governed by
the unperturbed one-electron Green's function at the surface of the metal
${\mathcal G}^{(0)}_{R,\sigma}(\omega-i\delta)$
and the scattering amplitude
$t_{\sigma}(\omega-i\delta)$ due to the impurity.
The latter, given by $\frac{\Delta}{\pi\rho_0} G_{d,\sigma}(\omega-i\delta)$
in the Anderson model \cite{Anderson,UKSZZ}, can be approximated as a sum of
three Lorentzian according to the scattering on the upper and lower d-levels
and the Kondo resonance, respectively. After a straightforward calculation 
the final expression obtained for the tunneling density of states is
\begin{equation}
\label{roveg}
  \delta\rho_{R}(\omega) =
     \frac{[\Im {\mathcal G}^{(0)}_{R}(\omega-i\delta)]^2}
        {\pi\rho _0}\,
       \biggl \{
  \frac{(q_{R}+\varepsilon)^2}{\varepsilon^2+1}- 1+ C_R
       \biggr \}
\end{equation}
where the spin index $\sigma$ was dropped, $q_{R}={\Re {\mathcal
    G}^{(0)}_{R}(\omega-i\delta)} /{\Im {\mathcal
    G}^{(0)}_{R}(\omega-i\delta)}$,
    $\varepsilon=(\omega-\varepsilon_K)/{T_K}$, $T_K$ is the Kondo
    temperature, $\ep_K$ is a shift with respect to the Fermi energy,
    and $\rho_0$ is the density of states at the Fermi level for one
    spin direction. $C_R$ arises from potential
    scattering on the d-level \cite{UKSZZ} and corresponds to a weakly
    energy dependent Friedel oscillation.

The first part of Eq.~(\ref{roveg}) coming from the scattering by the Kondo
resonance gives a Fano line shape in the tunneling LDOS, controlled by the
parameter $q_{R}$.  The fit on the experimental data for a Co atom on a Au
(111) surface \cite{Madhavan} gave excellent agreement with fitting parameters
being consistent with the predictions of an NCA calculation combined with band
structure results \cite{UKSZZ}. 

To calculate the distance dependence of $q_{R}$ and $C_{R}$, i.e., of the line
shape, the tunneling of electrons from the tip (1) into the 3-dimensional Au
bulk states as well as (2) into the 2-dimensional Au(111) surface band
\cite{Chen} was considered. In both cases a free electron-like band structure
was assumed \cite{UKSZZ}.  Whereas the periodic changes of the line shape
between Fano and Lorentzian ones and the decrease in the overall amplitude
with increasing distance were demonstrated for both cases 
(see Fig.~\ref{fstm} (b)), the
precise dependence of the line shape on $R$ is not reproduced by the
simplifying assumption of a free electron band structure \cite{UKSZZ}.  

There is a question concerning the role of the bulk and
surface states in the formation of the Kondo resonance and in the tunneling.
According to novel examinations \cite{abs1,abs2} it seems that the bulk states
have the dominating role in the former, and the surface states in the latter.

\section{Size dependence of the Kondo effect}

In the Kondo ground state the spin is completely screened by the
conduction electrons, thus the impurity spin must move together with a
compensation cloud in the conduction electron band and its scale is
the only relevant scale, namely, the Kondo coherence length $\xi
\sim v_F/T_K$ which can be extremely large for lower Kondo
temperatures, e.g. $\xi=10^4{\rm \AA}$ for $T_K=1$K. In the mesoscopic samples
that scale can be easily matched by the size of the sample and it was
conjectured \cite{Bergmann} that, in that case, the full spin
compensation cannot be developed, thus a suppression of the Kondo
effect could be expected. 

\begin{figure}
\begin{center}
\includegraphics[height=2.5cm]{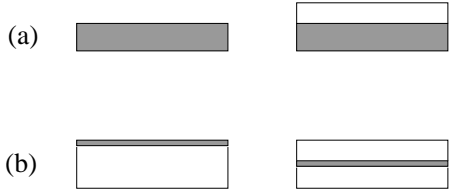}
\caption{The setups of experiments where the shadowed areas
    represent the dilute magnetic alloy and the clean ones the pure
    metal. The Kondo amplitude increases as (a) the alloys are covered by pure
    metal (b) the position of thin layer of the
    alloy is moving away from the surface.}
\label{fanis}
\end{center}
\end{figure}

In the 90's several experiments \cite{BG,exp11,exp12,exp2,Roth} were performed
on thin films and narrow wires of dilute magnetic alloys to measure the Kondo
compensation cloud. In most of the cases \cite{BG,exp11,exp12} a suppression
of the Kondo resistivity amplitude and no essential change in the Kondo
temperature were observed for small sample sizes. Covering a thin layer of
magnetic alloys by another pure metal layer (see Fig.~\ref{fanis} (a)), a
partial recovery of the Kondo signal was found \cite{prox,Blachly} which was
even smaller for more disordered overlayers \cite{proxdis}. These cannot be
attributed to the reduced size of the Kondo screening cloud due to the size of
the sample as only the energy separation of the metallic electron levels are
relevant.  As far as the level separation is smaller than $T_K$ only the shape
and size of the Kondo cloud is modified \cite{Affleck, Zarand}.

The experiments in ballistic samples were explained by the theory of the
spin-orbit-induced surface anisotropy \cite{UZGY,UZ1,UZ2,UZ3}. 
According to that theory a magnetic impurity in a metallic host with strong
spin-orbit interaction experiences a surface anisotropy having the following
form for flat surfaces:
\begin{equation}
  \label{eq:anisham}
  H=K_d ({\bf n}{\bf S})^2
\end{equation}
where ${\bf n}$ is the normal direction of the experienced surface element and
${\bf S}$ is the spin of the impurity, and $K_d>0$ is the anisotropy constant.
The most slowly decaying contribution is non-oscillating, and inversely
proportional to the distance $d$ measured from the surface, and is in the
range of $\frac{0.01}{(\textrm{d}/\textrm{\AA})}\,eV < K_d <
\frac{1}{(\textrm{d}/ \textrm{\AA})}\,eV$ \cite{UZGY,UZ1}.
\begin{figure}
\begin{center}
\includegraphics[height=2cm]{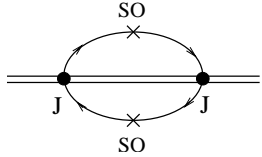}
\caption{The self-energy diagram generating the surface anisotropy 
for the impurity spin in leading order. The
double line represents the spin, the single one the conduction
electrons. The solid circles stand for the exchange interaction and the
$\times$ for the spin-orbit interaction on different host atoms.}
\label{fdiag}
\end{center}
\end{figure}

The surface contribution to the self-energy was calculated on basis of diagram
Fig.~\ref{fdiag} \cite{UZGY,UZ1} which is second order in the exchange coupling
and the spin-orbit (SO) scattering as well, both calculated for $l=2$
scattering.  For the spin-orbit scatterers an Anderson type model is taken
with spin-orbit interaction on the d-level of the scatterer and an average was
taken all over the sample \cite{UZGY,UZ1}.
\begin{figure}
\begin{center}
\includegraphics[height=4.5cm]{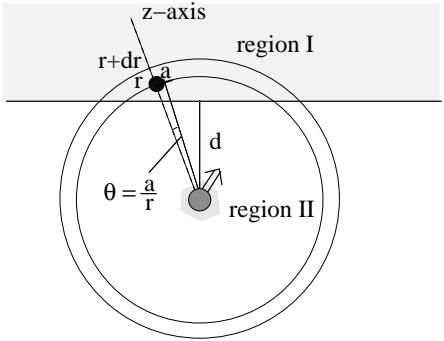}
\caption{The different regions of host atoms nearby the impurity. Only shells
  ($r$, $r+dr$) truncated by the surface contribute.}
\label{fmiss}
\end{center}
\end{figure}
The surprising $1/d$ dependence can be explained in the following
way. One of the two spin-orbit scatterers in diagram Fig.~\ref{fdiag}
must experience of the finite size by missing the scatterers in the
area indicated on the figure Fig.~\ref{fmiss} as region I, while the
other one can be very nearby the impurity (region II on
Fig.~\ref{fmiss}).  The number of the most effective missing
scatterers can be estimated as proportional to $d^3$, while of the second
type as the number of scatterers in the neighborhood of the spin which
is calculated to be between 100-1000. That large number appears as a
prefactor.  The transition concerning the angular momentum can be
treated in a frame, where the $z$ axis connects the certain scatterer
and the impurity. The magnetic quantum number changes e.g. as $m\to
m\pm1$. The overlap of those localized $d$-states with the conduction
electrons must be taken. The amplitude of the electron spherical wave
function centered at the impurity on an atomic scale $a$ ($a\ll d$)
nearby the axis is $1/r (a/r)^{|m|}$ at distance $r\sim d\gg a$.  $r^{-1}$
comes from the radial amplitude of a spherical wave, while $r^{-|m|}$
from the angular dependence ($Y_{lm}\sim (\sin\theta)^{|m|}$). Namely,
the scattering atom of size $a$ is seen from the spin inside the angle
$\theta\sim a/r$. Therefore the largest contribution is due to the
states $m=0$ and $m=\pm1$.  The overlap of the radial wave function
with the atomic orbital of the scatterer is, however, oscillating,
thus it contains an extra factor in the form $\sum_{n=0} (S^m_n\sin
k_F r (k_F r)^{-n} +C^m_n\cos k_F r (k_F r)^{-n} )$ (see Eq.s (13a-c)
in Ref.~\cite{UZ1}), where the amplitude $S^m_n$ and $C^m_n$ are zero
for even and odd $n$, respectively, and $k_F$ is the Fermi
wavelength. In the product of the ingoing and outgoing waves the
non-oscillating parts arise from terms proportional to $\sin^2 (k_F
r)$ or $\cos^2 (k_F r)$. In the case of combination $m=0$, $m=\pm 1$
waves the terms proportional to $C^{m=0}_{n=2} C^{m=\pm 1}_{n=2}$ and
$S^{m=0}_{n=1} S^{m=\pm 1}_{n=3}$ contribute and that results in an extra
$(k_F d)^{-1}$ factor. Thus the contribution from the missing
scatterers is $(d/a)^3 1/d^2 (a/d)^{|m|} (a/d)^{|m|\pm1}$. In that way the
largest contribution for $m=0, \pm 1$ is proportional to
$1/d$. Similar contribution arises when the incoming and outgoing
waves have identical $|m|=1$.

\begin{figure}[h]
\begin{center}
\includegraphics[height=3cm]{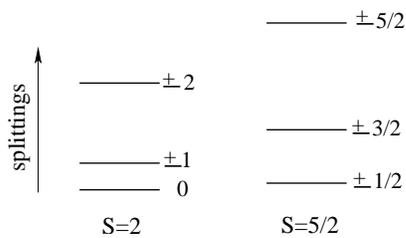}
\caption{The level splitting due to the surface anisotropy 
  for integer ($S=2$) and half-integer ($S=5/2$) spins.}
\label{fsplit}
\end{center}
\end{figure}

According to the anisotropy there are different splitting schemes for integer
and half-integer spins (see Fig.~\ref{fsplit}). For integer spins the ground
state is a singlet, whereas for half-integer spins it is a Kramer's doublet.
Thus for integer spins the anisotropy causes size effects \cite{UZ2} in
mesoscopic samples in agreement with the early experiments
\cite{BG,exp11,exp12,exp2,Roth}.  The surface anisotropy was tested later also
by new kind of experiments, measuring other spin dependent quantities (e.g.
the magnetoresistance \cite{G,BZ}, the thermopower \cite{Strunk}, and impurity
spin magnetization \cite{Se}), or in samples with half-integer spins
\cite{JG52,JG52yes,UZ3} which could be all explained with the theory of
surface anisotropy.

The thermopower of a homogeneous sample is zero even if the diameter
of the measured wire is changing along its length. In case of
surface anisotropy the percentage of blocked spins are higher in a
narrow region of the wire than in the broader one. Thus having
nominally the same impurity concentration, the concentration of the
effectively free spins are different, therefore measurable thermopower
occurs \cite{Strunk}.

The most exciting experiment probably was performed by Giordano and coworkers
\cite{JG} (see Fig.~\ref{fanis} (b)) where different multilayer structures
composed of Au and Au(Fe) films were examined positioning the overlayer only
on one side or on both sides of the film. The results gave a good agreement
with the theory of surface anisotropy also quantitatively.

\section{Dephasing and non-equilibrium transport}

The magnetic impurities play an important role in transport properties of
mesoscopic systems in cases of implanted impurities and contamination like
small amount of Fe impurities or oxidation of copper with forming copper oxide
($S=1$ \cite{Haesen}). 

\subsection{Dephasing}

The dephasing at low temperature has recently attracted considerable interest
and also intensified by the debate over the saturation at low temperature.
Magnetic impurities with degenerate spin states can always contribute to the
dephasing by spin flip processes. The non-elastic scattering rate has a
maximum at the Kondo temperature and freezes out at low temperatures where the
Kondo singlet ground state is formed, which do not exhibit dynamical feature.
The most appropriate study of that is a recent one, where the numerical
renormalization group has been applied \cite{ZBDA}. Such dephasing rate with
Fe doped samples has been observed by e.g. Haesendonck et al.
\cite{Haesendonck} showing maximum at $T_K$. The surface anisotropy can reduce
and even block the spin-flip processes for integer spin (e.g. Fe, Cr, CuO) if
the anisotropy constant $K_d$ is comparable to or dominating over the $T_K$,
respectively. On the other hand, for half-integer spin the anisotropy has only
minor role by freezing the spin into a degenerate Kramer's doublet and only
some prefactors can be modified. The dephasing is usually measured in
equilibrium situation as magnetoresistance
\cite{magexp1,magexp2,Natelson,AHexp}, universal conductance fluctuation
\cite{Mohanty}, and Aharonov-Bohm oscillation in mesoscopic rings in magnetic
fields \cite{AHexp}. In case of out-of-equilibrium situation where the
distribution of electrons is broadened by applying larger voltages on short
wires, dephasing becomes even possible if that broadening dominates the
anisotropy. Non-equilibrium measurements \cite{neq} were also carried out but
there were no direct evidences for the presence of magnetic impurities,
however, classical two-level systems with energy splitting can play a role
similar to the integer magnetic impurities with anisotropy.

The size dependence of dephasing was observed in PdAu nanowires
\cite{Natelson} with large spin-orbit scattering. As the diameters of the
wires were reduced the dephasing time increases. That is consistent with the
concept of surface anisotropy as in that case in smaller wires the integer
spins can be blocked.  That effect is just the opposite to the case where
extra dephasing scatterers are assumed at the surface, e.g. due to
copper-oxide.

\subsection{Energy relaxation}

\begin{figure}
\begin{center}
\includegraphics[height=3cm]{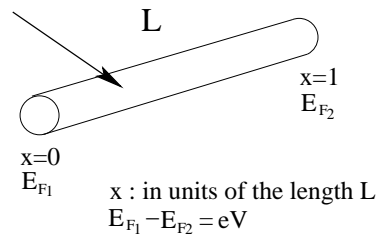}
\caption{The wire of length L. The arrow points to the position of the
attached tunnel junction.}
\label{fwire}
\end{center}
\end{figure}

The out-of-equilibrium transport in short wires with large bias voltage has
attracted great interest following the crucial experiments at Saclay
\cite{Saclay}. The distribution of electron energies were measured between the
contacts at different position by attaching extra tunnel junction to the side
of the wire (see Fig.~\ref{fwire}).
The wires were in the diffusive limit. The measured electron distribution in
many cases shows a step-like distribution due to the electrons coming and
accelerated from one of the contacts and decelerated from the other
one. Even at very low temperature the distribution, however, does not show
sharp steps but they are smeared due to a finite relaxation rate of the
electrons. Altshuler, Aronov et al. \cite{Alt} predicted an energy relaxation
rate due to the electron-electron scattering with $\ep^{-3/2}$ dependence on
the energy and well defined amplitude where the energy $\ep$ is measured from
the Fermi energy. 

Some of the experiments can be explained by that but in other cases
other relaxation mechanisms must exist. The Saclay group
phenomenologically introduced an extra interaction between the
electrons which has a $1/\ep^2$ singularity where $\ep$ is the energy
transfer. Several authors suggested \cite{KG,GG,K,Z,KZ,GGA} that the
energy exchange is mediated by Kondo impurities (magnetic and
structural defects) which was supported later also by experiments in
magnetic field \cite{Anthore}. It has been known since a long time
\cite{SZ} that such interaction is singular in the energy transfer and
recently Kaminski and Glazman \cite{KG} called the attention to
similar $1/\ep^2$ singularity suggested by the Saclay group. Using
that mechanism the electron transport is determined by the diffusive
Boltzmann equation and compared with the experimentally determined
electron distributions and the impurity concentrations were adjusted
\cite{GG,K,Z,KZ,GGA}.  In some cases the estimated magnetic impurity
concentrations using that method are much larger than those determined
from the dephasing rate, even by two orders of magnitude.

There are additional features. In the case of split Fermi energy due to the
bias, Kondo resonances appear at both Fermi energies, but as the steps in the
occupation numbers are reduced the Kondo temperature is also reduced assuming
that the bias voltage $e |V| < k T_K$. In those cases the weak coupling Kondo
theory (logarithmic approximation) can be applied. Furthermore, the
non-equilibrium distribution of the electrons makes possible spin flip
relaxation  by creating electron-hole pairs with almost zero energy like in
the Korringa relaxation processes. That relaxation rate is proportional to $e
|V|$ ($e |V|\gg k T$) and that results in an infrared cutoff smearing and even
blocking the Kondo effect. 

The effect of the spin-orbit-induced anisotropy on the energy
relaxation in mesoscopic wires was examined recently \cite{UJZ}.
There the non-equilibrium distribution function of a wire with length
$L$ (see Fig.~\ref{fwire}) was determined by solving the diffusive Boltzmann
equation
\begin{equation}
\begin{split}
    \frac{\partial f(\ep,x)}{\partial t}&-\frac{1}{\tau_D}
      \frac{\partial^2 f(\ep,x)}{\partial^2 x} + I_{\mathrm
        coll.}(\{f\})=0 \\
    I_{\mathrm coll.}(\{f\})&=\int dE
   \bigl \{f(\ep) [1-f(\ep-E)] W(\ep, E) \\
    &-[1-f(\ep)] f(\ep-E) W(\ep-E,-E) \bigr \}
\label{BE}
\end{split}
\end{equation}
taking into account in the scattering rate  due to magnetic
impurities ($W(\ep, E)$) the surface anisotropy, as well.  In Eq.~(\ref{BE})
$\tau_D=\frac{L^2}{D}$ is the diffusion constant, $f$ is assumed not
depending on the spin, and $x$ denotes the position in the wire in the
units of $L$.  Similar to the case of finite magnetic field \cite{GGA}
there are also first order processes contributing to the scattering
rate and the spin occupation numbers $p_M$s depend also on the
voltage.  Calculating the latter from the first order processes the
Boltzmann equation was solved self-consistently starting with the
solution without inelastic scattering mechanism $f^{(0)}(\ep,x)=(1-x)
n_F (\ep-\frac{e V}{2})+x n_F (\ep+\frac{e V}{2})$ and using the
following collision integral
\begin{equation}
\begin{split}\label{Icoll}
I^{(2)}&_{\mathrm coll.}(\{f\})=\int dE\int d\ep'
    K^{S}_{MM'} (E,\ep,\ep', K_d) \times \\
&\times\bigl \{p_{M} f(\ep) f(\ep') [1-f(\ep-E)]\cdot \\
  &\hspace*{5em} \cdot [1-f(\ep'+E+K_d M^2-K_d M'^2)] \\
  &\hspace*{1em} -p_{M'} [1-f(\ep)] [1-f(\ep')] f(\ep-E)\cdot \\
&\hspace*{5em} \cdot f(\ep'+E+K_d M^2 - K_d M'^2)\bigr \}
\end{split}
\end{equation}
where the kernel $K^S_{MM'}$ describes electron-electron interaction mediated
by Kondo impurities with surface anisotropy.
\begin{figure}[!h]
\centerline{\includegraphics[height=1.7cm]{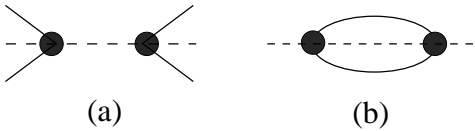}}
\caption{\label{fig2} The diagrams used for calculating (a) the kernel and (b)
  the Korringa lifetime of the impurity spin. The
  solid lines denote the conduction electrons, the dotted lines the impurity
  spin, and the blob is the Kondo coupling.}
\end{figure}
For simplicity, the case $S=1$ (i.e. for CuO in Cu wires \cite{Haesen}) was
considered and an appropriate constant value ${\tilde J}$ was used
instead of the renormalized Kondo couplings depending on $M,M'$
\cite{UJZ}.  The kernel $K^S_{MM'}$ was calculated in the Kondo model
with anisotropy \cite{UZ2} according to the diagram on Fig.~\ref{fig2}
(a).  As the weak dependence on the Korringa lifetime \cite{Korringa}
$\tau_K$ of the impurity spin the value for $K_d=0$ according to
diagram Fig.~\ref{fig2} (b) was used \cite{UJZ}. At each step of the
self-consistent numerical calculation, both the spin occupation
numbers and the Korringa lifetime were updated from the actual $f$.

\begin{figure}[!h]
\begin{center}
\includegraphics[height=5.5cm]{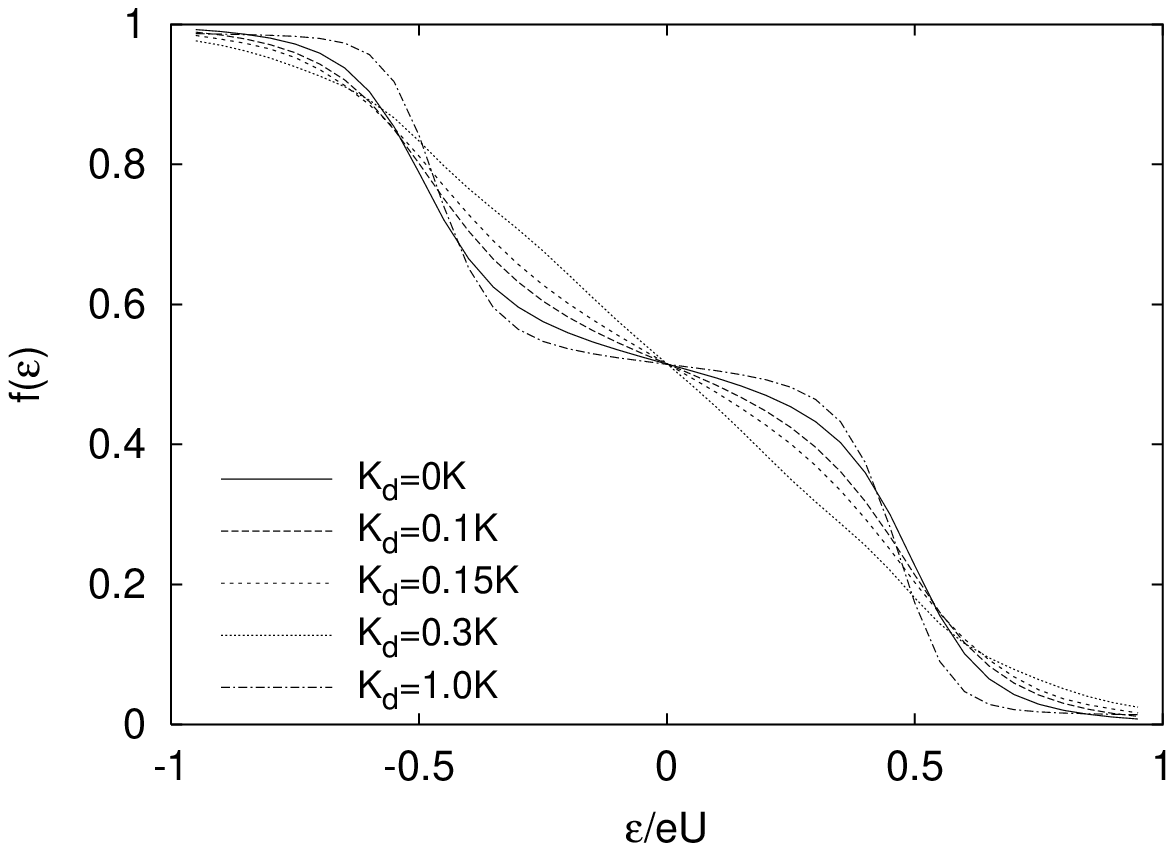}
\includegraphics[height=5.5cm]{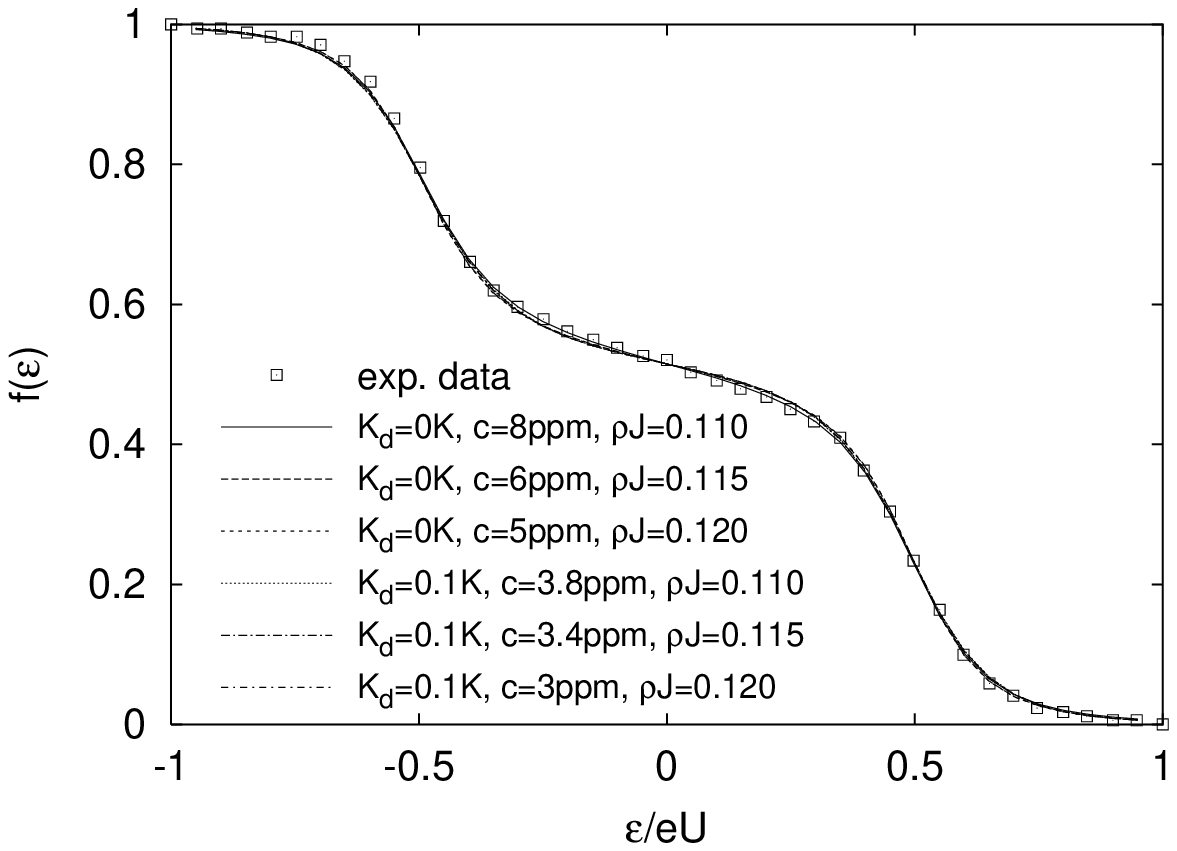}
\caption{\label{fig34} The calculated distribution function at
  $x=0.485$ for different values of $K_d$. The other parameters used
  are $c=8$ppm $\rho_0 {\tilde J}=0.11$, $U=0.1$mV, and
  $\tau_D=2.8$ns.  (b) Fit on the experimental data of Cu wires
  \cite{Anthore} for different $K_d$, $\rho_0 {\tilde J}$, and $c$
  triads. The other parameters used $U=0.1$mV and $\tau_D=2.8$ns.}
\end{center}
\end{figure}

In Fig.~\ref{fig34} (a) the dependence of the distribution function on
the strength of the anisotropy is illustrated.
Increasing $K_d$ first the
energy transfer is getting larger but for larger $K_d$ the ground state is
frozen in, similar to the magnetic field dependence discussed in
Ref.~\cite{GGA}. We can conclude
that the contribution of magnetic impurities is enhanced
or unchanged in case when for the strength of the anisotropy $K_d<e U$.
For $K_d\sim 0.1-0.2$K which is a
good estimation for the strength of the anisotropy for the wires with width of
$\sim 45$nm and thickness of $\sim 85-110$nm used in the experiments, the
energy relaxation is only slightly affected by the anisotropy.

In Fig.~\ref{fig34} (b) fits on the experimental data of Cu wires
\cite{Anthore} can be seen for different $K_d$, $\rho_0 {\tilde J}$,
and $c$ triads. We can see that equally good fit can be achieved with
different $K_d$, $\rho_0 {\tilde J}$ (corresponding to the Kondo
temperature) and $c$ triads. That uncertainty is further increased if a
distribution in $K_d$ is also included. At fix $\rho_0 {\tilde J}$, the larger
the $c$ is the smaller the necessary $K_d$.

The half-integer case must be very similar to the case without surface
anisotropy because of the degeneracy in the ground state, and only the
spin dependent prefactors are different.

Thus the role of surface anisotropy in electron dephasing and energy 
relaxation is very different for integer and half-integer spins 
having a singlet and a Kramer's doublet ground state,
respectively. In the first case for low temperature and thermal equilibrium
the spin dynamics and therefore the dephasing are frozen out while in the
out-of equilibrium metallic wire experiments that can reenter. That suggests a
pronounced size dependence and very different concentration for the
dynamically active impurities in the dephasing and the out-of-equilibrium
wire experiments. In the case of half-integer spin that cannot be expected.
Further careful experiments for the size dependence and implanted impurities
are required.

\section{Point contacts with magnetic impurities}

\begin{figure}[h]
\begin{center}
\includegraphics[height=2.5cm]{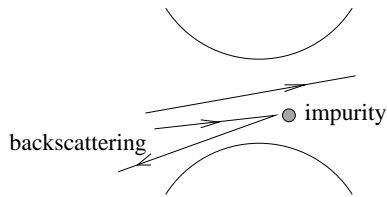}
\caption{The dominating backscattering process by the magnetic
impurity in the PC.}
\label{fpc}
\end{center}
\end{figure}

In point contact (PC) the magnetic impurities in the contact region cause
very strong electron scattering in the energy region of $T_K$ around
the Fermi energy. Thus they strongly influences the I-V
characteristics at zero bias. The reduction in the current flowing
through the contact is due to the electron backscattering (see
Fig.~\ref{fpc}) thus at
zero bias there is resistivity increase. A thorough study of the  
Kondo effect in ultra small  CuMn PCs has been  
carried out \cite{Yanson1,Yanson2}. Rather surprisingly, in this case not a  
suppression but an \emph{orders of magnitude increase} of   
the Kondo temperature has been reported.  

As shown in Ref.~\cite{ZarandUdvardi1,ZarandUdvardi2}, these anomalies can be
well explained by the presence of LDOS fluctuations. For a small PC, even a
weak channel quantization induces huge LDOS fluctuations
\cite{ZarandUdvardi1,ZarandUdvardi2} which become larger and larger with
decreasing contact sizes.  $T_K$ depends on the inverse of the LDOS
exponentially ($T_K\sim exp(-2\rho_0 J)$). The LDOS has a weak dependence on
the energy, thus the change at the Fermi energy is dominating, that may
produce an extremely wide distribution of the Kondo temperatures for
impurities in the contact region.  The zero bias anomaly of the PC, however,
turns out to be dominated by impurities with the largest $T_K$, while those
with very small $T_K$ do not show up at larger voltages. Indeed, in
Ref.~\cite{ZarandUdvardi1,ZarandUdvardi2} the effect of these fluctuations was
taken into account through a modified renormalization procedure, and a perfect
agreement was found between the calculated and experimentally determined
anomalous amplitude of the Kondo signal.

It was also predicted by the theory \cite{ZarandUdvardi1,ZarandUdvardi2} that
this effect should be much less pronounced for alloys with large $T_K$ as
$T_K$ is less sensitive to the change in $\varrho_0$ in that case. That has
been later confirmed by the experiments studying Cu(Fe) alloys \cite{A}.

\section{Orbital Kondo effect due to dynamical defects}

In many different experiments typical Kondo logarithmic temperature
dependence (see Ref.~\cite{Cox}) has been observed and it was argued
that is not due to the electron spin.  It has been suggested many
years ago \cite{Cox} that dynamical defects where an atom is sitting in
a double well potential can be described by a quasispin and the
electron-defect interaction depends on that quasispin and also on the
angular momenta of the electron. The Hamiltonian is very similar to
the spin Kondo Hamiltonian \cite{Kondo}. In that model the dominating
processes are the electron screening and electron assisted tunneling
of the atom between the two wells. The corresponding couplings are
non-commuting, thus logarithmic terms are expected. The variables in
the couplings are of orbital origin, while the real electron spin is a
silent conserving one. In that case if the region below the Kondo
temperature could be reached, the ground state has a complicated
structure due to the spin degeneracy. The main consequence is that the
Fermi liquid behavior at $T<T_K$ is replaced by non-Fermi liquid
behavior showing $T^{1/2}$ temperature dependence.

Several experiments have been reported where the observed low
temperature anomalies were attributed to TLS Kondo defects
\cite{PbGeTe,disloc,RalphBuhrman,Upad,Keijsers,Zarcomm}, namely to
\emph{dynamical} structural defects: they disappear under annealing and
did not, or only slightly depended on magnetic field.

There is an easy way to distinguish between slow TLS and Kondo-like
scatterers \cite{Halbritter}. As it has been discussed concerning the
point contact with magnetic impurities, the dominating dynamical
processes are the back-scatterings of the electrons resulting in an
increase of resistivity. In case of slow, classical TLS at low
temperature the electron must have large enough energy to excite the
TLS, thus that processes have a lower threshold voltage
different for different scatters. Thus the resistivity increases with
the voltage. On the other hand, as it was discussed earlier in the case
of Kondo scattering, the resistivity maximum is at zero bias.

The most of zero-bias anomalies \cite{Halbritter} are surprisingly Kondo-like.
A logarithmic increase of the resistivity attributed to the presence of
dislocations or substitutional tunneling impurities has been observed in
various systems \cite{PbGeTe,disloc}.  The most spectacular experiments were
carried out in Cu and Ti PCs where a two-channel Kondo-like $\sim\sqrt{T}$ and
$\sqrt{V}$ non-Fermi liquid scaling behavior due to non-magnetic scatterers
has been observed in the dynamical resistance \cite{RalphBuhrman,Upad},
however the nature of the scatterers are still unknown. The widths of the
zero-bias anomalies were associated with the Kondo temperature, $T_K \sim 5K$.
In another beautiful experiment a repeated switching of the zero-bias anomaly
between two curves in the presence of some slow TLSs has been observed in
amorphous PCs \cite{Keijsers}, which could be consistently explained assuming
that a slow fluctuator influences the splitting of one or two fast Kondo TLSs
close to it \cite{Zarcomm} and that causes the change in the zero-bias
anomalies.  

There is a further set of experiments \cite{balkashin} where an
alternating voltage was superimposed on a constant bias $V_0$, $V(t) = V_0 +
V_1 \cos(\omega t)$. As far as the characteristic frequency (e.g. Kondo
temperature) of the mechanism responsible for the zero-bias anomaly is large
compared to $\hbar \omega$, the measured $I-V$ characteristic is just the time
average of the current: $\langle{I(t)}\rangle = I (V_0) + \frac{1}{4}
\left(\frac{\partial^2 I}{\partial V^2}\right)_{V = V_0} V_1^2$.

The experiments agree with that, but those are far from being
conclusive. In case of slow TLS when the actual voltage is in the range of the
TLS energy $\Delta$ rectification occurs as current starts to flow when the
actual voltage exceeds $\Delta$ \cite{Halbritter}. That is also frequency
independent. Thus the earlier expectation that in case of slow TLS
strong frequency dependence is expected is not correct. 

\begin{figure}[htbp]
  \centering
  \includegraphics[height=2.5cm]{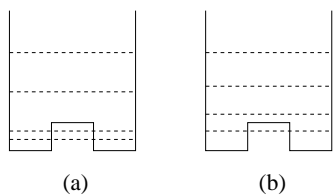}
  \caption{Tower with the energy levels of the atomic motion;
    (a) two levels below the barrier (b) only one level below the
    barrier.}
  \label{FT}
\end{figure}

The model known as orbital two-channel Kondo model was strongly
criticized by Kagan and his coworkers \cite{Kagano} and later by
Altshuler et al. \cite{Altcrit}. Kagan's argument was that the
tunneling rate of the atom between the two wells is drastically
reduced as the atom is moving with the electron screening cloud and
logarithmic terms survive only in much lower energy region than the electronic
band with cutoff. The actual range is determined
roughly by the third excited state of the atom in the well. That
argument has recently made more transparent \cite{Altcrit} using a
simplified model where the atom is moving in an infinitely tall
potential tower with a double potential at the bottom (see
Fig.~\ref{FT}). Thus the atom does not become free well above the
Debye temperature.

That model was treated numerically in the logarithmic approximation, and it
was shown that including more than the first six levels the Kondo temperature
is drastically reduced due to the reduction in the tunneling rate. That
problem can be avoided by assuming that the second level is just above the
barrier \cite{Bord} (see Fig.~\ref{FT}), thus tunneling does not occur. The
question remained open, whether the Kondo temperature can be large enough to
explain zero-bias anomalies in the energy range of $10$K and also that the
renormalized energy separation $\Delta$ between the two lowest levels can
allow the scaling to reach the temperature range $T<T_K$. Assuming an
intermediate atomic mass as 50 proton mass, the Kondo temperature is in the
range of $0.2$ K for very strong coupling, but the role of the energy
separation is far from being clear. In this way the observed large zero-bias
anomalies cannot be explained, but dephasing and energy relaxation rate can be
influenced. The experimental situation concerning the role of structural
defects in dephasing in metallic glasses and strongly disordered systems is
still subject of debates \cite{Mohanty,Lin1,Lin2}.

As in the tower model the relevant energy scale is the level spacing,
that and also the Kondo temperature can be increased by considering light
atoms like H which can be present in atomic form or e.g. as
water. Then the requested $T_K$ ($T_K\sim 10$K) can be reached and also 
the coupling is also strong, because the phase shift of the electron
scattering is nearby $\delta=\pi/2$ according to the Friedel sum
rule. The energy separation $\Delta$ should be studied by numerical
renormalization group valid even in that very strong coupling region.
Recent experiment \cite{CsonkaHalb} in Hungary strongly indicates the
role of atomic hydrogen in Pd and Pt contacts where the H content is
systematically changed.

\section{Conclusions}

In the coming years the identification of Kondo scatterers in the phenomena of
dephasing and energy relaxation will be a very important task. Theoretically
the possibility of the structural Kondo scatterers must be clarified. These
problems are very closely related and they should be studied considering
different samples containing different impurities and prepared in different
ways.

\section*{Acknowledgements}
  
  We are grateful to N. O Birge, V. M. Fomin, H. Grabert, J. Kroha, H.
  Pothier, and G. Zar\'and for useful discussions.  This work was
  supported by Hungarian grants OTKA F043465, T046303, T034243,
  TS040878, T038162, and grant No. RTN2-2001-00440.

\end{document}